# Best conventional solutions to the King's Problem

P.K.Aravind
Physics Department
Worcester Polytechnic Institute
Worcester, MA 01609
(Email: paravind@wpi.edu)

ABSTRACT

In the King's Problem, a physicist is asked to prepare a $d$-state quantum system in any state of her choosing and give it to a king who measures one of $(d+1)$ sets of mutually unbiased observables on it. The physicist is then allowed to make a control measurement on the system, following which the king reveals which set of observables he measured and challenges the physicist to predict correctly all the eigenvalues he found. This paper obtains an upper bound on the physicist's probability of success at this task if she is allowed to make measurements only on the system itself (the "conventional" solution) and not on the system as well as any ancillary systems it may have been coupled to in the preparation phase, as in the perfect solutions proposed recently. An optimal conventional solution, with a success probability of 0.7, is constructed in $d=4$; this is to be contrasted with the success probability of 0.902 for the optimal conventional solution in $d=2$. The gap between the best conventional solution and the perfect solution grows quite rapidly with increasing $d$.

Key words: quantum state retrodiction, quantum algorithms

## 1. Introduction

In the *King's Problem*, as it was termed in [1], a physicist is asked to prepare a $d$-state quantum system in any state of her choosing and give it to a king who measures one of $d+1$ sets of mutually unbiased observables on it. The physicist is then allowed to make a control measurement on the system, as well as any other systems it may have been coupled to in the preparation phase, following which the king reveals which set of observables he measured and challenges the physicist to predict correctly all the eigenvalues he found. This problem was solved in dimension $d=2$ in [2], following which some variants of it were treated in [3]-[5]. Solutions to the problem were then given in all prime dimensions [6] and in all prime power dimensions [7]. An experimental realization of the problem in $d=2$ was recently reported in [8].

The King's Problem is an example of a problem in quantum state retrodiction, in which the object is to determine the state of a quantum system at some time in the past on the basis of measurements made on it before and after that time. The solutions to the King's Problem mentioned in the previous paragraph were all obtained by allowing the physicist to couple the system to an ancillary system (or "ancilla") and make a control measurement on both the system and the ancilla. The physicist is able to pull off her trick by preparing a suitably entangled state of the system and ancilla and choosing a control measurement that allows her to rule out all but one of the permitted combinations of eigenvalues for each of the observable sets the king is allowed to measure, thereby allowing her to predict his results correctly when he finally reveals his choice of measurement. It should be stressed that the physicist succeeds at a limited task, namely, that of predicting (or retrodicting) what the king found in the one measurement he did actually make, and that her apparent knowledge of the outcomes of the other measurements has no connection with anything that actually transpired or might have transpired. A failure to appreciate this subtle (and not at all obvious) point could leave a spectator with the erroneous impression that the physicist has succeeded in determining sharp eigenvalues for all of several sets of mutually incompatible observables, in violation of one of the basic tenets of the quantum theory.

It is interesting to ask how unexpected (or "magical") the perfect solutions to the King's Problem are. One way of gauging this is to ask how well the physicist can do if she is not allowed to make use of an ancilla (and hence any entanglement) and is restricted to making measurements on the system itself. This method of attack is what is referred to in the title of this paper as the "conventional solution" to the King's Problem. It was shown in [8] that the best conventional solution to the King's Problem in $d=2$ has a success rate of 90.24%, which falls about 10% short of the perfect solution. The purpose of this paper is to generalize this analysis to arbitrary dimension $d$ and determine how close the best conventional solution comes to the perfect solution. The gap between the best conventional and perfect solutions is one measure of the crucial role of entanglement in this problem, and may also serve as a rough indicator of the degree of surprise one might justifiably expect to experience upon being shown this trick.

We demonstrate in this paper that the best conventional solution to the King's Problem in dimension $d$ has a probability of success, $P(d)$, that is bounded from above in the manner described by the inequality

$$P(d) \leq \frac{2\sqrt{d} + d - 1}{\sqrt{d}(1+d)} . \qquad (1)$$

For $d = 2$ the upper bound in (1) reduces to that found in [8]. Table 1 shows numerical values of the upper bound in (1) for several values of $d$, and it is seen that the gap between the upper bound and unity widens with increasing $d$, just as one would expect. This confirms the increasing efficacy of entanglement in obtaining a perfect solution to the King's Problem at higher $d$, and lends interest to an experimental realization of such a solution for $d \geq 3$ (with $d = 4$ suggesting itself as a particularly favorable case).

It might be asked whether the upper bound in (1) can ever actually be achieved. Ref.[8] presents a protocol that achieves this bound for $d = 2$, while we present a protocol in this paper that achieves it for $d = 4$. It can be shown by a finitary analysis that the bound cannot be achieved for $d = 3$, and we have not so far found a way to achieve it for any other values of $d$.

This paper is organized as follows. Section 2 lays out the argument leading to (1), with the proof of a key assertion being relegated to the Appendix. Section 3 shows the impossibility of achieving the upper bound in (1) for $d = 3$ and then goes on to construct a protocol that does achieve it for $d = 4$. Finally, Section 4 discusses a variant of the $d = 2$ problem in which the king is allowed to measure the spin of his qubit along any of the body diagonals of a cube and shows that there is a noticeable gap (of about 7%) between the best ancilla-assisted and conventional solutions in this case.

**2. Derivation of the upper bound (1)**

Before stating the King's Problem more precisely, we need to define what is meant by a set of mutually unbiased bases (or observables). The orthonormal bases $|\alpha_i\rangle$ and $|\beta_i\rangle$ ($i = 1, 2, ..., d$) of a $d$-state system are said to be mutually unbiased if $|\langle \alpha_i | \beta_j \rangle|^2 = 1/d$ for all $i$ and $j$. The term "unbiased" refers to the fact that any state of one basis is equally likely to yield any state of the other if a measurement in the latter basis is carried out on it. A set of bases is said to be mutually unbiased if all pairs of bases within it are mutually unbiased. It has been shown [9] that a $d$-state system has at most $(d+1)$ mutually unbiased bases, and an explicit prescription for constructing such bases has been given in all prime $d$ [1,10] as well as in all prime power $d$ [9,11,12]. The states comprising a basis can always be regarded as the simultaneous eigenstates of a complete sets of commuting observables whose eigenvalues provide a unique set of state labels for each state in the basis. The various sets of commuting observables, whose eigenstates are the mutually unbiased bases, are also spoken of as being mutually unbiased. Thus the notion of mutual unbiasedness applies equally to bases as it does to the sets of commuting observables that give rise to these bases.

The King's Problem, and its solution by the conventional method (which is the object of interest in this paper), can now be stated as consisting of the following steps:

(A) The physicist prepares a $d$-state quantum system in a state of her choosing and gives it to the king.
(B) The king measures a complete set of commuting observables on the system, and notes the eigenvalues of each of the observables he measures. The set he measures is one of the $(d+1)$ sets of mutually unbiased observables for the system.
(C) The physicist carries out a control measurement on the system. (In this problem there is no coupling of the system to an ancilla, so the measurement is made on nothing larger than the system itself).
(D) The king reveals which set of commuting observables he measured.
(E) The physicist is required to predict correctly the eigenvalues of all the observables measured by the king.

The solution to the problem requires specifying the physicist's strategy in steps (A) and (C) and then calculating her success probability at step (E); the optimal solution is the one that leads to the largest possible success probability at step (E). What state should the physicist prepare in step (A)? She could prepare an arbitrary pure or mixed state, but a moment's reflection shows that a mixed state can be ruled out. To see this, note that any mixed state can be expressed as a convex combination of an optimal pure state and other pure states that are certainly no superior to it. The loss of information involved in passing from the optimal pure state to a mixture, combined with the less than optimal character of the other pure states involved, will certainly lead to a decrease in the success probability for a mixture as compared to a pure state. It in fact turns out that the optimum pure state is an eigenstate of one of the (mutually unbiased) sets of observables measured by the king. This fact is proved in the Appendix, and we proceed here with this particular choice of preparation state to establish the bound (1).

The physicist therefore prepares the system in an eigenstate of one of the unbiased sets of observables and gives it to the king. If the king measures this particular set of observables, which he does with a probability of $1/(d+1)$, the physicist can predict his results perfectly, and so she need only worry about designing a strategy that would allow her to maximize her chances of success if he should happen to pick one of the other sets of observables. The way she does this is by measuring an observable whose $d$ nondegenerate eigenstates each have a small (ideally zero) overlap with all but one state in each of the $d$ bases other than the preparation basis. When she measures this observable in step (C), only the eigenstates having a non-negligible overlap with the state found by the king are likely to be returned, and so, if she adopts the strategy of picking the single state within each basis that has a non-negligible overlap with the returned eigenstate, she is highly likely to succeed.

A hint on picking the right control measurement in step (C) is provided by an examination of the optimal conventional solution in the $d = 2$ case [8]. There the physicist prepares the system (a qubit) in a spin up state along the z-axis, thereby guessing the king's result perfectly if he measures along this direction, and chooses her control measurement to be midway between the x- and y-axes. This "equiangular" choice of measurement basis might lead one to suspect that a similar equiangular choice (i.e. a basis that straddles the remaining $d$ unbiased bases symmetrically) might also lead to an optimal solution in the $d$-dimensional case. A detailed analysis shows that the idea of an equiangular basis can be used as a convenient crutch to establish the bound (1), and that this bound can be realized only if the equiangular basis actually exists.

We now proceed to substantiate these claims. Let $|\psi_j^i\rangle$ ($i = 0,1,..,d$ and $j = 1,..,d$) denote the $j$-th state of the $i$-th unbiased basis. Suppose the physicist prepares the system in one of the states of the 0-th basis and subsequently carries out her control measurement in the orthonormal basis $|\chi_k\rangle$ ($k = 1,..,d$). The way in which she uses her control measurement is as follows. Each state $|\psi_j^i\rangle$ is associated with one of the measurement states $|\chi_k\rangle$, namely the state with which it has the maximum overlap (if a state has equal overlaps with two or more measurement states, it is arbitrarily assigned to any one of them). If, at the end of this assignment, each measurement state $|\chi_k\rangle$ has exactly one state from each unbiased basis associated with it, the measurement basis will be said to be "well-conditioned" and, if not, "ill-conditioned". A well-conditioned basis allows the physicist to make a unique prediction for the state found by the king in each unbiased basis, but an ill-conditioned basis does not. The way one gets around the hurdle posed by an ill-conditioned basis is simple: one reassigns some of the basis states $|\psi_j^i\rangle$ to other measurement states in such a way that the measurement basis becomes well-conditioned (at the cost, of course, of lowering the detection probability of certain of the unbiased basis states). We will assume henceforth that any control measurement basis is always rendered well-conditioned before it is used.

Let $f(i,j)$ be the probability that the state $|\psi_j^i\rangle$ obtained by the king as a result of his measurement is identified correctly by the physicist; if $|\chi_k\rangle$ is the measurement state that $|\psi_j^i\rangle$ is associated with, then $f(i,j) = |\langle \chi_k | \psi_j^i \rangle|^2$. The physicist's success probability in guessing the king's result at step (E) is then

$$P(d) = \frac{1}{d+1} \cdot 1 + \sum_{i=1}^{d} \sum_{j=1}^{d} \frac{1}{d+1} \cdot \frac{1}{d} \cdot f(i,j), \qquad (2)$$

which is a weighted sum of the success probabilities if the king measures in the preparation basis (first term) or in one of the other bases (second term), it being assumed that he picks a basis totally at random. Let $F(k)$ denote the sum of all the $f(i,j)$ associated with a particular measurement state $|\chi_k\rangle$. Then the terms within the double summation in (2) can be regrouped according to measurement state to yield

$$P(d) = \frac{1}{d+1}\left[1 + \frac{1}{d}\sum_{k=1}^{d} F(k)\right], \qquad (3)$$

where each $F(k)$ is the sum of $d$ of the $f(i,j)$.

The success probability (3) achieves its maximum value if each $F(k)$ achieves its maximum value. We now show that $F(k) \leq pd$, where $p = (\sqrt{d} + d - 1)/d\sqrt{d}$, which, when substituted into (3), yields the bound (1). To show that $F(k) \leq pd$, we consider the problem of constructing a measurement state $|\chi_k\rangle$ that leads to the maximum value of $F(k)$. The state $|\chi_k\rangle$ can be conveniently expressed as a superposition of all the states $|\psi_j^i\rangle$ for which it acts as the "signal". Suppressing the basis label $k$ for brevity, we can write

$$|\chi\rangle = \sum_{i=1}^{d} a_i e^{i\phi_i} |\psi^i\rangle, \qquad (4)$$

where the subscript $j$ on $|\psi_j^i\rangle$, which is a function of both the measurement label $k$ and the unbiased basis label $i$, has also been suppressed. Any measurement state can always be expressed in this form, where the $a_i$ and $\phi_i$ are real parameters. The quantities $a_i$ and $\phi_i$ are to be determined from the requirement that $F = \sum_{i=1}^{d} |\langle \psi^i | \chi \rangle|^2$ is maximized subject to the normalization constraint $\langle \chi | \chi \rangle = 1$. This problem can be attacked using the method of Lagrange multipliers and leads to a rather complicated set of equations involving not only the $a_i$ and $\phi_i$ but also the phases $\theta_{jl}$ arising from the overlaps $\langle \psi^j | \psi^l \rangle = e^{i\theta_{jl}} d^{-1/2}$ between pairs of basis states. However a way around this impasse is provided by seeking a maximum of $F$ under the more relaxed conditions that the $\theta_{jl}$ are not fixed but can vary freely, since such a maximum clearly provides an upper bound on the true maximum. Maximizing $F$ with respect to the $a_i, \phi_i$ and $\theta_{jl}$ leads to a set of equations whose solution is readily seen to be $\phi_i = 0, \theta_{jl} = 0$ (i.e. all phases vanishing) and all the amplitudes $a_i$ equal to each other. The single amplitude $a$ can then be fixed from the normalization condition $\langle \chi | \chi \rangle = 1$ and leads to the maximum value $pd$ for $F$, where $p = (\sqrt{d} + d - 1)/d\sqrt{d}$. This proves the inequality $F \leq pd$, and hence the bound (1).

A discussion of why the choice of an eigenstate (as the preparation state) in the above derivation is optimal can be found in the Appendix.

### 3. Optimal conventional solution for $d = 4$.

The proof of the previous section shows that the upper bound in (1) can be achieved if it proves possible to construct a set of measurement states $|\chi_k\rangle$ having the following properties:

P1. Each state $|\chi_k\rangle$ can be expressed as an equally weighted linear combination of $d$ states $|\psi_j^i\rangle$, with exactly one state coming from each unbiased basis other than the physicist's preparation

basis. In other words, each measurement state $|\chi\rangle$ (we drop the subscript $k$) should be expressible in the form

$$|\chi\rangle = N \sum_{i=1}^{d} e^{i\phi_i} |\psi^i\rangle, \qquad (5)$$

where $N$ is a normalization constant and subscripts have been omitted from the $\psi^i$.

P2. The phases $\phi^i$ in (5) can be chosen so that $\langle \psi^i | \chi \rangle = N\left(1 + \frac{d-1}{\sqrt{d}}\right)$ for all $i$. This implies, from (5), that $N = \left[d + \sqrt{d}(d-1)\right]^{-1/2}$ and hence that $|\langle \psi^i | \chi \rangle|^2 = \frac{\sqrt{d} + d - 1}{d\sqrt{d}}$ for all $i$.

P3. There exist $d$ orthonormal measurement states having properties P1 and P2.

The property P2 is particularly difficult to satisfy because one has only $d$ phases $\phi^i$ with which to cancel out the $d(d-1)/2$ phases arising from the inner products of the various $|\psi^i\rangle$. This phase cancellation can always be achieved for a single $\langle \psi^i | \chi \rangle$, but then one has no freedom left to engineer this cancellation for other values of $i$. Even if P2 can somehow be satisfied by a single measurement state, it is far from clear that one can construct $d$ orthonormal states having this property, as required by P3. Thus the task of constructing a set of measurement states satisfying properties P1-P3 would appear to be impossible in general.

An investigation shows this pessimism to be justified for $d = 3$. One begins by trying to construct a state of the form (5) that satisfies property P2. States of the form (5) can be constructed in $3^3 = 27$ different ways by picking one state from each of the three bases available for the purpose. For each of these forms of the state (5), one picks the phases $\phi^i$ so that $\langle \psi^i | \chi \rangle$ has the required value for a single value of $i$, but then one finds that the remaining $\langle \psi^i | \chi \rangle$ do not have this value, thereby signaling the failure of this construction.

For $d = 4$, however, the above construction succeeds and leads to a solution that achieves the upper bound in (1). To obtain this solution, we note that a four-state system can be realized by a pair of qubits and that such a system possesses the five mutually unbiased sets of observables (and bases) shown in Table 2. The physicist again begins by preparing a pair of qubits in one of the states of basis 0 and giving it to the king, who measures any one of the five sets of mutually unbiased observables (shown in the first column of Table 2) on it. Our task is now to design a set of measurement states that satisfy conditions P1-P3. To this end, we consider states of the form

$$|\chi\rangle = \frac{1}{\sqrt{10}}\left[|\psi_i^1\rangle + b|\psi_j^2\rangle + c|\psi_k^3\rangle + d|\psi_l^4\rangle\right] \qquad (6)$$

which is just (5) specialized to $d = 4$ but with the state subscripts restored and $b, c$ and $d$ standing for complex numbers of modulus unity. For all $4^4 = 256$ choices of $i, j, k$ and $l$ we have to investigate whether it is possible to choose $b, c$ and $d$ so that $|\langle \psi_i^1 | \chi \rangle|^2 = |\langle \psi_j^2 | \chi \rangle|^2 = |\langle \psi_k^3 | \chi \rangle|^2 = |\langle \psi_l^4 | \chi \rangle|^2 = 5/8$, as required by condition P2. We find that there are 32 choices of $i, j, k$ and $l$ that meet this condition. These choices, together with the corresponding values of $b, c$ and $d$, are shown in Table 3. It now remains to see if one can pick four mutually orthogonal states out of this set of 32 that the physicist can use as her measurement basis. It turns out that one can actually pick 32 such bases, and they are shown in Table 4. Any one of these bases can be used by the physicist as her control measurement basis. The expression (6) not only indicates the makeup of each measurement state in terms of the states of the different unbiased bases, but also encodes instructions on how the physicist is to respond if she obtains the state $i, j, k, l$ as a result of her control measurement: she should guess the state $i$ in basis 1, $j$ in basis 2, $k$ in basis 3 or $l$ in basis 4. Her success probability with this strategy is 0.7.

**4. A variation of the VAA problem in $d = 2$**

In the original VAA version of the King's Problem [2], the king is given a qubit and allowed to measure its spin along any one of three orthogonal directions. We now consider a slight variation of this problem in which the king is allowed to measure the spin along any of the four body diagonals of a cube. We show how the VAA basis of Ref. [2] can be used to obtain a solution to the problem that, while not perfect, is appreciably better than the best conventional solution that can be obtained to the same problem.

The trick begins with the physicist preparing a pair of qubits in the entangled state $|\psi\rangle = (|00\rangle + |11\rangle)/\sqrt{2}$, and then giving the first (object) qubit to the king and retaining the second (ancilla) qubit in her possession. The king is allowed to measure the spin of the object qubit along any one of the four body diagonals of a cube. We take the cube to be oriented so that the unit vectors from its center to its eight vertices are

$$\pm \hat{n}_1 = \pm \frac{1}{\sqrt{3}}(1,1,1), \quad \pm \hat{n}_2 = \pm \frac{1}{\sqrt{3}}(-1,1,1), \quad \pm \hat{n}_3 = \pm \frac{1}{\sqrt{3}}(-1,-1,1) \text{ and } \pm \hat{n}_4 = \pm \frac{1}{\sqrt{3}}(1,-1,1). \quad (7)$$

Let $|\hat{x}, \hat{y}\rangle$ denote a state of the system and ancilla qubits in which the spin of the former is up along $\hat{x}$ and that of the latter is up along $\hat{y}$. It is easily verified that the entangled state $|\psi\rangle$ can be expressed in four alternative ways as

$$|\psi\rangle = (|\hat{n}_1, \hat{n}_4\rangle + |-\hat{n}_1, -\hat{n}_4\rangle)/\sqrt{2}$$
$$= (|\hat{n}_2, \hat{n}_3\rangle + |-\hat{n}_2, -\hat{n}_3\rangle)/\sqrt{2}$$
$$= (|\hat{n}_3, \hat{n}_2\rangle + |-\hat{n}_3, -\hat{n}_2\rangle)/\sqrt{2} \quad , \quad (8)$$
$$= (|\hat{n}_4, \hat{n}_1\rangle + |-\hat{n}_4, -\hat{n}_1\rangle)/\sqrt{2}$$

from which it follows that when the king measures the spin of the object qubit along a particular diagonal he finds it to be up or down with equal likelihood and also collapses the ancilla qubit into a spin up state that is the reflection of the object qubit's state in the x-z plane.

For her control measurement, the physicist chooses an observable whose nondegenerate eigenstates are the four VAA basis states [2]

$$|\chi_1\rangle = \frac{1}{\sqrt{2}}|00\rangle + \frac{1}{2}e^{i\pi/4}|01\rangle + \frac{1}{2}e^{-i\pi/4}|10\rangle; \quad (\hat{n}_4, \hat{n}_1)$$
$$|\chi_2\rangle = \frac{1}{\sqrt{2}}|00\rangle - \frac{1}{2}e^{i\pi/4}|01\rangle - \frac{1}{2}e^{-i\pi/4}|10\rangle; \quad (\hat{n}_2, \hat{n}_3)$$
$$|\chi_3\rangle = \frac{1}{2}e^{-i\pi/4}|01\rangle + \frac{1}{2}e^{i\pi/4}|10\rangle + \frac{1}{\sqrt{2}}|11\rangle; \quad (-\hat{n}_3, -\hat{n}_2) \quad (9)$$
$$|\chi_4\rangle = -\frac{1}{2}e^{-i\pi/4}|01\rangle - \frac{1}{2}e^{i\pi/4}|10\rangle + \frac{1}{\sqrt{2}}|11\rangle; \quad (-\hat{n}_1, -\hat{n}_4)$$

which have the property that the object and ancilla qubits are significantly polarized along the cube body diagonals related to each other by reflection in the x-z plane (the pair of diagonals for each state is indicated in parentheses after it). A straightforward calculation of the overlaps of the VAA states with the various object-ancilla states that can result from the king's measurement reveals the results shown in Table 5. One sees that each VAA state has a practically vanishing overlap with one state in each of the four bases picked by the king, thus allowing the physicist to predict what result the king most likely got if he measured along each of the four body diagonals. For example, if she gets the state $|\chi_1\rangle$, she will predict that the king found spin up along $\hat{n}_1$ or spin down along $\hat{n}_2$ or spin up along $\hat{n}_3$ or spin up along $\hat{n}_4$. One also sees, by looking across any horizontal row of Table 5, that any object-ancilla state produced by the king activates the wrong VAA state with a probability of only 0.069, thereby allowing the physicist to predict the king's result correctly with a probability of 0.933 if she follows this method.

This can be contrasted with the best conventional solution to the problem. For this purpose, the physicist prepares the qubit in a spin up state along the diagonal $\hat{n}_1$, thus guaranteeing a perfect result if the king measures along this direction, and performs her control measurement along a direction that maximizes her chances of guessing the king's result right if he measures along one of the other directions. A detailed examination of all directions on the Bloch sphere shows that the optimum direction for the control measurement makes an angle of $\arctan(-4\sqrt{2}) \approx 100°$ with $\hat{n}_1$ and is inclined away from it along the great circle arc joining it to $\hat{n}_3$. Her success probability with this choice of control measurement is $(15 + \sqrt{33})/24 = .864$, which is significantly lower than what can be achieved with the modified VAA method.

## 5. Concluding remarks

This paper obtains an upper bound, given in Eqn.(1), on the success probability of a conventional solution to the King's Problem which may help to clarify the degree of surprise inherent in the perfect solutions to this problem found recently in prime [6] and prime power [7] dimensions. The gulf between the conventional and perfect solutions widens with increasing dimension, just as one would expect, and demonstrates the increasing efficacy of entanglement in securing the perfect solution at higher dimensions. An optimal conventional solution has been presented in $d = 4$, with a success rate of 70%. An experimental realization of the King's Problem in $d = 4$ would be very welcome, since it would provide a more dramatic confirmation of the efficacy of entanglement than in the $d = 2$ case. The technology for carrying out such an experiment certainly seems to be in place [13]. On the theoretical front, it would be interesting to settle the question of whether the bound in Eqn.(1) can be achieved for dimensions other than 2 and 4.

**Acknowledgement.** I would like to thank Berge Englert for his useful comments on an earlier draft of the paper and particularly for pointing out a gap in the proof of (1), which has now been closed.

## APPENDIX

We wish to show that using an arbitrary (pure) state as the preparation state, rather than an eigenstate as in Sec.2, does not lead to an increase in the upper bound in (1). The most general strategy that the physicist can use in conjunction with an arbitrary preparation state is the following: she can make educated guesses if the king measures in $r$ of the bases, and use her control measurement to infer his result only if he measures in the remaining $s \equiv d + 1 - r$ bases. We denote her success probability $P(d,r) = P_g(d,r) + P_c(d,s)$, where the first and second terms are the parts coming from the guesses and control measurement, respectively. It is not difficult to show that $P_g(d,r)$ is maximized by choosing a preparation state that has equal overlaps with one state from each of the $r$ "guess" bases, with these states then being chosen as the king's result in these bases. To see this, write the preparation state as $|\Psi\rangle = \sum_{i=1}^{r} b_i |\psi^i\rangle$, where the $|\psi^i\rangle$ are the guessed states in these bases and the $b_i$ are complex amplitudes. We need to maximize the sum of the guess probabilities, $\sum_{i=1}^{r} |\langle \psi^i | \Psi \rangle|^2$, subject to the normalization constraint $\langle \Psi | \Psi \rangle = 1$. An upper bound on this maximum can be obtained by taking all the $b_i$ to be real and equal to each other, with their common value being determined by the normalization constraint. This leads to the result

$$P_g(d,r) \leq \frac{\sqrt{d}+r-1}{\sqrt{d}(d+1)} \qquad (A1)$$

We next turn to the task of obtaining an upper bound on $P_c(d,s)$. For her control measurement, the physicist measures an observable whose $d$ nondegenerate eigenstates provide clues to the king's result if he measures in one of the $s$ bases excluding the preparation and guess bases. Let $|\chi_k\rangle$ $(k=1,..,d)$ denote the eigenstates of the physicist's control observable (we will refer to these as "measurement states", as in Sec.2). As before, we will ensure that the measurement states are well-conditioned i.e. that each $|\chi_k\rangle$ serves as a "signal" for exactly one state from each of the $s$ bases in question. In an obvious extension of (4), one can write a typical measurement state as

$$|\chi\rangle = \sum_{i=1}^{s} a_i e^{i\phi_i} |\psi^i\rangle, \qquad (A2)$$

with the same notational conventions as in (4). The expression for the success probability now changes from (2) to

$$P_c(d,s) = \sum_{i=1}^{s}\sum_{j=1}^{d} \frac{1}{d+1} \cdot p(i,j) \cdot f(i,j) \qquad (A3)$$

where $f(i,j) = |\langle \chi_k|\psi_j^i\rangle|^2$ as before and $p(i,j)$ is the probability that the king obtains the $j-$th state in the $i-$th basis (of the $s$ bases in question) when he carries out his measurement. The terms in (A3) can be regrouped according to the measurement states they are associated with and written as

$$P_c(d,s) = \frac{1}{d+1}\sum_{k=1}^{d}\sum_{i=1}^{s} p(i)|\langle \chi_k|\psi^i\rangle|^2 \equiv \frac{1}{d+1}\sum_{k=1}^{d} F(k), \qquad (A4)$$

where we have omitted the label $j$ from $p(i)$ and $|\psi^i\rangle$ because it plays no essential role in what follows (note that $j$ is completely determined by $k$ and $i$ anyway). To maximize the success probability (A4), we need to maximize $F(k)$ for each $k$. This problem is the same for each $k$, so we can look at the maximization of $F = \sum_{i=1}^{s} p(i)|\langle \chi|\psi^i\rangle|^2$ with $|\chi\rangle$ given by (A2) and subject to the normalization constraint $\langle \chi|\chi\rangle = 1$. As in Sec.2, we can tackle this problem with the aid of Lagrange multipliers and get an upper bound on $F$ by taking all the phases $\phi_i$ and $\theta_{jl}$ to be equal to zero. A maximization with respect to the amplitudes $a_i$ then leads to the equations

$$2\sqrt{d}\left[p(i)+\lambda\right]a_i + \left[2p(i)+\lambda\right]\sum_{j\neq i} a_j = 0, \quad i=1...s, \qquad (A5)$$

where $\lambda$ is a Lagrange multiplier and the sum over $j$ in the second term ranges over all integers from 1 to $s$ with the exception of $i$.

The equations (A5) are a set of homogeneous linear equations for the amplitudes $a_i$, and so possess a non-trivial solution only if the determinant of the coefficient matrix vanishes. This condition fixes the Lagrange multiplier $\lambda$, and the amplitudes $a_i$ can then be solved for. However, even without a detailed solution, it is obvious that the $a_i$ are of the form

$$a_i = f\left(p(i), \{p(j)|j\neq i\}\right), \quad i=1,...s, \qquad (A6)$$

where $f$ is the *same* function for all the $a_i$, and the arguments of $f$ have been divided into two groups to emphasize the different ways in which they contribute to $f$. It is obvious that if (A6) is substituted back into $F$, a totally symmetric function of all the probabilities $p(i)$ results. The maximization of $F$ over the $p(i)$ then leads to a preparation state for which all the $p(i)$ are the same. Not only is the value of $p(i)$ independent of $i$, but it is also independent of $k$ because the maximization of $F$ is independent of $k$. The value of $p(i)$ can be fixed from the fact that it is the same for all states in an unbiased basis, and that the sum of these probabilities is unity, from which it follows that $p(i)=1/d$. In other words, the optimum initial state is one that yields any state in any of the $s$ unbiased bases with a probability of $1/d$. The equality of the $p(i)$ implies the equality of the $a(i)$, with their common value being fixed at $\left[(d+1)(\sqrt{d}+1)\right]^{-1/2}$ by the normalization constraint $\langle\chi|\chi\rangle=1$. Calculating $F$ under these conditions and putting it back into (A4) yields the upper bound

$$P_c(d,s) \leq \frac{\sqrt{d}+s-1}{(d+1)\sqrt{d}}. \qquad (A7)$$

We are now in a position to calculate the total success probability $P(d,r)$. For $r=d+1$ we can calculate it from (A1) alone, while for $r=0$ we can calculate it from (A7) alone with $s=d+1$; in both these cases we get the same upper bound,

$$P(d,0) = P(d,d+1) \leq \frac{1+\sqrt{d}}{1+d}. \qquad (A8)$$

For all other values of $r$ we must sum the expressions (A1) and (A7), and we find that

$$P(d,r) \leq \frac{2\sqrt{d}+d-1}{\sqrt{d}(1+d)}, \quad r=1,...,d. \qquad (A9)$$

Note that the upper bound in (A9) is independent of $r$ and is greater than that in (A8) for all $d \geq 2$. Note also that the upper bound in (A9) is identical to that found in Sec.2, thus proving our claim that an eigenstate is at least as good as any other choice of initial preparation state.

In those cases in which the upper bound in $P(d,1)$ can be achieved by a suitable choice of control measurement observable, there is an alternative to choosing the preparation state to be an eigenstate and proceeding as in Sec.2: one can choose the preparation state to be one of the "measurement" states (i.e. eigenstates of the control measurement observable) and perform the control measurement in the basis from which the preparation state was earlier picked; this realizes the upper bound in $P(d,d)$ and is the complement of the earlier procedure in that the roles of the preparation and measurement bases are exchanged, as are the numbers of bases for which guesses are made and for which the control measurement is used (I am indebted to Berge Englert for this remark).

| $d$ | 2 | 3 | 4 | 5 | 8 | 9 |
|---|---|---|---|---|---|---|
| Max $P(d)$ | .9024 | .7887 | .7000 | .6315 | .4972 | .4667 |

TABLE 1. The upper bound on the success probability, as predicted by Eqn.(1), for several values of the dimension $d$.

| | | | | |
|---|---|---|---|---|
| Z1,1Z | $\lvert\psi_1^0\rangle = 1000$ | $\lvert\psi_2^0\rangle = 0100$ | $\lvert\psi_3^0\rangle = 0010$ | $\lvert\psi_4^0\rangle = 0001$ |
| X1,1X | $\lvert\psi_1^1\rangle = 1111$ | $\lvert\psi_2^1\rangle = 1\bar{1}1\bar{1}$ | $\lvert\psi_3^1\rangle = 11\bar{1}\bar{1}$ | $\lvert\psi_4^1\rangle = 1\bar{1}\bar{1}1$ |
| Y1,1Y | $\lvert\psi_1^2\rangle = 1ii\bar{1}$ | $\lvert\psi_2^2\rangle = 1\bar{i}i1$ | $\lvert\psi_3^2\rangle = 1i\bar{i}1$ | $\lvert\psi_4^2\rangle = 1\bar{i}\,\bar{i}\,\bar{1}$ |
| XY,YZ | $\lvert\psi_1^3\rangle = 1\bar{1}ii$ | $\lvert\psi_2^3\rangle = 11\bar{i}i$ | $\lvert\psi_3^3\rangle = 11i\bar{i}$ | $\lvert\psi_4^3\rangle = 1\bar{1}\,\bar{i}\,\bar{i}$ |
| YX,ZY | $\lvert\psi_4^1\rangle = 1i\bar{1}i$ | $\lvert\psi_2^4\rangle = 1\bar{i}1i$ | $\lvert\psi_3^4\rangle = 1i1\bar{i}$ | $\lvert\psi_4^4\rangle = 1\bar{i}\,\bar{1}\,\bar{i}$ |

TABLE 2. Mutually unbiased sets of observables/bases for a system of two qubits. Each row shows a pair of commuting observables for a system of two qubits (written as products of the Pauli and identity operators of the individual qubits), followed by their four simultaneous eigenstates arranged according to the eigenvalue signatures $++, +-, -+$ and $--$. Each eigenstate is written $\lvert\psi_j^i\rangle$, with the superscript indicating its basis (0-4) and the subscript its position within the basis (1-4). The shorthand notation $\lvert\psi_j^i\rangle = abcd$ is used to indicate that $\lvert\psi_j^i\rangle$ has the (unnormalized) form $a\lvert 00\rangle + b\lvert 01\rangle + c\lvert 10\rangle + d\lvert 11\rangle$ with respect to the standard basis of the pair of qubits (with $i = \sqrt{-1}$ and a bar over a number indicating its negative). The observables, as well as the bases, in the different rows of the table are mutually unbiased.

| # | i | j | k | l | b | c | d | # | i | j | k | l | b | c | d |
|---|---|---|---|---|---|---|---|---|---|---|---|---|---|---|---|
| 1 | 1 | 1 | 1 | 1 | $-i$ | $-i$ | $-i$ | 17 | 3 | 1 | 3 | 2 | 1 | 1 | $i$ |
| 2 | 1 | 1 | 2 | 2 | $-i$ | 1 | 1 | 18 | 3 | 1 | 4 | 1 | 1 | $-i$ | 1 |
| 3 | 1 | 2 | 3 | 1 | 1 | 1 | $-i$ | 19 | 3 | 2 | 1 | 2 | $i$ | $i$ | $i$ |
| 4 | 1 | 2 | 4 | 2 | 1 | $i$ | 1 | 20 | 3 | 2 | 2 | 1 | $i$ | 1 | 1 |
| 5 | 1 | 3 | 1 | 3 | 1 | $-i$ | 1 | 21 | 3 | 3 | 3 | 4 | $-i$ | 1 | 1 |
| 6 | 1 | 3 | 2 | 4 | 1 | 1 | $i$ | 22 | 3 | 3 | 4 | 3 | $-i$ | $-i$ | $-i$ |
| 7 | 1 | 4 | 3 | 3 | $i$ | 1 | 1 | 23 | 3 | 4 | 1 | 4 | 1 | $i$ | 1 |
| 8 | 1 | 4 | 4 | 4 | $i$ | $i$ | $i$ | 24 | 3 | 4 | 2 | 3 | 1 | 1 | $-i$ |
| 9 | 2 | 1 | 1 | 4 | 1 | 1 | $-i$ | 25 | 4 | 1 | 3 | 3 | $i$ | $i$ | $i$ |
| 10 | 2 | 1 | 2 | 3 | 1 | $i$ | 1 | 26 | 4 | 1 | 4 | 4 | $i$ | 1 | 1 |
| 11 | 2 | 2 | 3 | 4 | $-i$ | $-i$ | $-i$ | 27 | 4 | 2 | 1 | 3 | 1 | 1 | $i$ |
| 12 | 2 | 2 | 4 | 3 | $-i$ | 1 | 1 | 28 | 4 | 2 | 2 | 4 | 1 | $-i$ | 1 |
| 13 | 2 | 3 | 1 | 2 | $i$ | 1 | 1 | 29 | 4 | 3 | 3 | 1 | 1 | $i$ | 1 |
| 14 | 2 | 3 | 2 | 1 | $i$ | $i$ | $i$ | 30 | 4 | 3 | 4 | 2 | 1 | 1 | $-i$ |
| 15 | 2 | 4 | 3 | 2 | 1 | $-i$ | 1 | 31 | 4 | 4 | 1 | 1 | $-i$ | 1 | 1 |
| 16 | 2 | 4 | 4 | 1 | 1 | 1 | $i$ | 32 | 4 | 4 | 2 | 2 | $-i$ | $-i$ | $-i$ |

TABLE 3. The 32 measurement states, for a system of two qubits, that satisfy conditions P1-P3 of the text. Each of the states has the form in Eqn.(6), with the values of $i, j, k, l, b, c$ and $d$ indicated.

| 1 | 1, 11, 22, 32 | 2 | 1, 11, 24, 30 | 3 | 1, 12, 21, 32 | 4 | 1, 15, 22, 28 |
|---|---|---|---|---|---|---|---|
| 5 | 2, 11, 22, 31 | 6 | 2, 12, 21, 31 | 7 | 2, 12, 23, 29 | 8 | 2, 16, 21, 27 |
| 9 | 3, 9, 22, 32 | 10 | 3, 9, 24, 30 | 11 | 3, 10, 23, 30 | 12 | 3, 13, 24, 26 |
| 13 | 4, 9, 24, 29 | 14 | 4, 10, 21, 31 | 15 | 4, 10, 23, 29 | 16 | 4, 14, 23, 25 |
| 17 | 5, 11, 18, 32 | 18 | 5, 15, 18, 28 | 19 | 5, 15, 20, 26 | 20 | 5, 16, 17, 28 |
| 21 | 6, 12, 17, 31 | 22 | 6, 15, 18, 27 | 23 | 6, 16, 17, 27 | 24 | 6, 16, 19, 25 |
| 25 | 7, 9, 20, 30 | 26 | 7, 13, 18, 28 | 27 | 7, 13, 20, 26 | 28 | 7, 14, 19, 26 |
| 29 | 8, 10, 19, 29 | 30 | 8, 13, 20, 25 | 31 | 8, 14, 17, 27 | 32 | 8, 14, 19, 25 |

TABLE 4. The 32 measurement bases that lead to the optimal conventional solution to the King's Problem in $d = 4$. The states in the various bases are just the ones in Table 3, and are identified by the same numbers as there (note that each state occurs in exactly 4 bases). Any one of these bases can be used by the physicist in her control measurement to solve the King's Problem with a success rate of 70%.

|  | $\lvert\chi_1\rangle$ | $\lvert\chi_2\rangle$ | $\lvert\chi_3\rangle$ | $\lvert\chi_4\rangle$ |
| --- | --- | --- | --- | --- |
| $\lvert\hat{n}_1,\hat{n}_4\rangle$ | 0.311 | 0.311 | 0.311 | 0.0669 |
| $\lvert-\hat{n}_1,-\hat{n}_4\rangle$ | 0.0223 | 0.0223 | 0.0223 | 0.933 |
| $\lvert\hat{n}_2,\hat{n}_3\rangle$ | 0.0223 | 0.933 | 0.0223 | 0.0223 |
| $\lvert-\hat{n}_2,-\hat{n}_3\rangle$ | 0.311 | 0.0669 | 0.311 | 0.311 |
| $\lvert\hat{n}_3,\hat{n}_2\rangle$ | 0.311 | 0.311 | 0.0669 | 0.311 |
| $\lvert-\hat{n}_3,-\hat{n}_2\rangle$ | 0.0223 | 0.0223 | 0.933 | 0.0223 |
| $\lvert\hat{n}_4,\hat{n}_1\rangle$ | 0.933 | 0.0223 | 0.0223 | 0.0223 |
| $\lvert-\hat{n}_4,-\hat{n}_1\rangle$ | 0.0669 | 0.311 | 0.311 | 0.311 |

TABLE 5. Each entry in the table is the squared overlap of the VAA state at the head of its column with the object-ancilla state at the beginning of its row. Note that each VAA state has a practically vanishing overlap with one state of each of the four measurement bases, thus allowing each VAA state to be used to make a prediction about the outcome of the king's measurement in each of the four bases. The numerical entries in the table arise from the trigonometric quantities $\frac{1}{2}\cos^4\frac{\theta}{2}=0.311, \frac{1}{2}\sin^4\frac{\theta}{2}=0.0223, \frac{3}{2}\cos^4\frac{\theta}{2}=0.933$ and $\frac{3}{2}\sin^4\frac{\theta}{2}=0.0669,$ where $\theta=\cos^{-1}\left(1/\sqrt{3}\right)$ is the angle between a body diagonal and a fourfold axis of the cube.